\documentclass[epj]{svjour}

\usepackage{graphicx,color}
\usepackage{amsmath}
\usepackage{epstopdf}

\newcommand{\D}{\mathrm{d}}
\newcommand{\e}{\mathrm{e}}
\newcommand{\half}{\frac{1}{2}}
\newcommand{\be}{\begin{equation}}
\newcommand{\ee}{\end{equation}}
\newcommand{\bea}{\begin{eqnarray}}
\newcommand{\eea}{\end{eqnarray}}
\newcommand{\ba} {\begin{align} }
\newcommand{\ea} {\end{align} }
\newcommand{\eps}{\epsilon}

\newcommand{\kbt}{k_{\mathrm{B}}T}

\newcommand{\lb}{l_{{\rm B}}}
\newcommand{\ld}{\lambda_{{\rm D}}}

\newcommand{\lgc}{l_{{\rm GC}}}

\newcommand{\ra}[1]{\textcolor{black}{#1} } 

\begin{document}


\title{Osmotic pressure between arbitrarily charged planar surfaces: a revisited approach}

\author{Ram M. Adar, David Andelman}
\institute{Raymond and Beverly Sackler School of Physics and Astronomy\\ Tel Aviv
University, Ramat Aviv, Tel Aviv 69978, Israel}

\abstract{The properties of ionic solutions between charged surfaces are often studied within the Poisson-Boltzmann framework, by finding the electrostatic potential profile. For example, the osmotic pressure between two charged planar surfaces can be evaluated by solving coupled equations for the electrostatic potential and osmotic pressure. Such a solution relies on symmetry arguments and is restricted to either equally or oppositely charged surfaces. Here, we provide a different and more efficient scheme to derive the osmotic pressure straight-forwardly, without the need to find the electrostatic potential profile. We derive analytical expressions for the osmotic pressure in terms of the inter-surface separation, salt concentration, and arbitrary boundary conditions. Such results should be useful in force measurement setups, where the force is measured between two differently prepared surfaces, or between two surfaces held at a fixed potential difference. The proposed method can be systematically used for generalized Poisson-Boltzmann theories in planar geometries, as is demonstrated for the sterically modified Poisson-Boltzmann theory. 
     }

\maketitle

\section{Introduction}
\label{sec1}

Ionic solutions are governed by the interplay between electrostatic interactions and the ion mixing entropy~\cite{Israelachvily,VO,Safinyabook}. Within mean field theory (MFT), this interplay results in the Poisson-Boltzmann (PB) equation, which determines the spatial variation of the electrostatic potential. For 1:1 monovalent salt, the PB equation in Gaussian units reads
\be
\label{eq1}
\frac{\eps}{4\pi}\nabla^2\Psi=2 n_b e \sinh \left(\frac{e \Psi}{\kbt}\right),
\ee
where $\Psi$ is the electrostatic potential, $\eps$ the dielectric constant of the solution, $n_b$  the bulk concentration of ions, $e$ the elementary charge, and $\kbt$ is the thermal energy. \ra{Note that within PB theory, ions are assumed to be point-like and uncorrelated, corresponding to rather dilute solutions.}

Consider a planar electrostatic setup, where two infinitely large, flat surfaces are aligned along the $x\,y$-plane and bound an ionic solution. \ra{Generally, surface charges are distributed heterogeneously~\cite{Patchy}, but as a convenient approximation, the surface-charge density can be modeled as constant.} Between two such homogeneously charged surfaces, the electrostatic potential, $\Psi(Z)$, varies along the normal direction to the bounding surfaces, taken here to be the $Z$ coordinate. Equation~(\ref{eq1}) can then be integrated analytically once, leading to
\be
\label{eq2}
-\frac{\eps}{8\pi}\Psi'^2+2 n_b \kbt \cosh \left(\frac{e \Psi}{\kbt}\right)=P,
\ee
where $\Psi'=\D\Psi/\D Z$. The integration constant, $P$, is the pressure in the electrolyte solution~\cite{Dan09,Maggs16}, and is a constant throughout the solution. The first term on the left-hand-side of Eq.~(\ref{eq2}) is the electrostatic contribution to the pressure, as given by the Maxwell stress tensor, and the second term is the van 't Hoff ideal gas contribution.

Evaluating Eq.~(\ref{eq2}) at the charged surface bounding the ionic solution leads to a relation between the surface charge, surface potential and pressure, known as the Grahame equation. On a broader scope, it is related to the contact theorem~\cite{Safinyabook,Grahame47} that exceeds the validity of the PB theory and is considered to be exact~\cite{Henderson81,Evans99,Dean03} \ra{for systems that are translationally invariant  in the $x\,y$-plane.}
For a single surface in contact with the electrolyte solution, the electrostatic potential and electric field decay to zero far away from the surface. Evaluating the pressure at such large distances yields the bulk van 't Hoff term,  $P_b=2n_b\,\kbt$.
On the other hand, when two surfaces bound an electrolyte solution, the bulk conditions are not met within the finite system, leading to a pressure difference between the bounded solution and the bulk one, $P_{\rm osm}=P-P_b$. This pressure difference is the {\it osmotic pressure}, $P_{\rm osm}$, and it depends on the inter-surface separation, $D$, and the electrostatic boundary conditions.

In this work, we consider two types of electrostatic boundary conditions. (i)~Constant potential (CP) --- surfaces held at a constant potential correspond to the Dirichlet boundary condition,
\bea
\label{CP}
\left.\Psi\right|_{Z=-D/2}&=&\Psi_{1}\, ,\nonumber\\
& & \nonumber\\
\left.\Psi\right|_{Z=D/2}~ &=&\Psi_{2}\, .
\eea
(ii)~Constant charge (CC) --- surfaces with fixed surface-charge densities, $\sigma_{1,2}$, correspond to the Neumann boundary condition via Gauss' law,
\bea
\label{CC}
 \left.\frac{\D \Psi}{\D Z}\right|_{Z=- D/2} &=& -\frac{4\pi}{\eps}\sigma_{1}\, , \nonumber\\
& & \nonumber\\
\left.\frac{\D \Psi}{\D Z}\right|_{Z= D/2}  ~~&=& ~ \frac{4\pi}{\eps}\sigma_{2} \, .
\eea
Here we assumed that the electric field does not penetrate the surfaces, as is often done in models due to a high dielectric mismatch between $\eps\approx80$ for water, and the value of the external medium.

For convenience, we rescale hereafter the important variables to be dimensionless:
\bea
\psi&\equiv& e\Psi/\kbt, \nonumber\\
p&\equiv &P/\left(2n_b\kbt\right)\, , ~~ \Pi\equiv P_{\rm osm}/\left(2n_b\kbt\right),
\label{eq5t}
\eea
where $\psi$, $p$ and $\Pi$ are, respectively, the dimensionless electrostatic potential, pressure and osmotic pressure.
Furthermore, the $Z$ coordinate and the inter-surface separation, $D$, are rescaled with the Debye screening length $\ld=1/\sqrt{8\pi \lb n_b}$, where $\lb=e^2/\left(\eps\kbt\right)$ is the Bjerrum length, such that
\be
z\equiv Z/\ld, ~~~  d\equiv D/\ld,
\ee
and the rescaled (dimensionless) electric field is
\be
\label{eq7t}
{\cal E}\equiv -\frac{\D\psi}{\D z}.
\ee
With these variables, Eq.~(\ref{eq2}) reduces to $p=-{\cal E}^2/2+\cosh\psi$, and the bulk value of $p$ reduces to unity. Therefore, the dimensionless osmotic pressure is,
\be
\label{eq3}
\Pi=p-1= -\half {\cal E}^2+2\sinh^2\frac{\psi}{2}.
\ee

Equation~(\ref{eq3}) can be used to derive an integral expression for $z(\psi)$, in an arbitrary interval $[z_1,z_2]$ in between the two surfaces
\be
\label{eq4}
z_2-z_1=\pm\int_{\psi_1}^{\psi_2}\frac{\D \psi}{\sqrt{4\sinh^2\frac{\psi}{2}-2\Pi}},
\ee
where $\psi_1=\psi(z_1)$ and $\psi_2=\psi(z_2)$, and the $\pm$ sign is a result of taking the square root of Eq.~(\ref{eq3}). Its sign depends on whether $\psi(z_2)<\psi(z_1)$ or vice versa, and is chosen to ensure that the entire right-hand-side of Eq.~(\ref{eq4}) is positive in order to match $z_2-z_1>0$ on the left-hand-side. Such $\pm$ signs appear in numerous equations throughout this work.

It is important to note that the field ${\cal E}=-\D\psi/\D z$ must maintain the same sign in the interval $z_1\le z \le z_2$, such that $\psi=\psi(z)$ is a monotonic function, which can be inverted as $z=z(\psi)$.  In addition, since the osmotic pressure is not known {\it a priori}, $\Pi$ should be viewed as a parameter in Eq.~(\ref{eq4}). Solving this equation yields a potential profile $\psi=\psi(z;\Pi)$ that depends on $\Pi$. The value of $\Pi$ is determined as to satisfy Eq.~(\ref{eq3}) for the given boundary conditions at $z=\pm d/2$.

So far we reviewed the standard scheme to obtain the osmotic pressure, $\Pi$. This procedure is useful, but has two disadvantages. First, it requires finding the full electrostatic potential profile, $\psi(z)$, as an intermediate step. Second, it is limited for the study of symmetric  or antisymmetric boundary conditions. For symmetric boundary conditions, where both surfaces have the same  fixed surface-charge density or potential, the electric field vanishes by symmetry in the mid-plane between the surfaces. Similarly, for antisymmetric boundary conditions of opposite surface charge or potential~\cite{Meier04,Safran05}, the potential vanishes in the mid-plane. As a consequence, the potential in both the symmetric and antisymmetric cases
is monotonic between the mid-plane and any of the two surfaces: $\left[-d/2,0\right]$ or $\left[0,d/2\right]$. On the other hand, for a non-monotonic potential, integrals of the form of Eq.~(\ref{eq4}) must be written separately for each interval where $\psi$ is monotonic. Due to such extra integrals, it is not possible to directly relate the osmotic pressure, inter-surface separation and the boundary conditions.

Below we revisit this problem and present an improved scheme for calculating the pressure between two charged surfaces, without the need to find the potential $\psi(z)$ and to ensure its monotonicity. The advantage is that the scheme enables us to solve the PB equation between any two arbitrary and asymmetric charged boundary conditions. Furthermore, we extend the same treatment to augmented PB theories such as the sterically modified one, provided that we restrict the system to have a planar symmetry.

\section{The Scheme}
\label{sec2}

As the pressure is determined by the boundary conditions and the separation $d$  between the surfaces, it can be found by extending the range of the integral in Eq.~(\ref{eq4}), such that $z_1$ and $z_2$ correspond to the positions of the two bounding surfaces. Then, $d=z_2-z_1$ and $\psi_1=\psi(z_1)$ and $\psi_2=\psi(z_2)$ are given by the boundary conditions at $z_{1,2}=\pm d/2$. When $\psi$ is non-monotonic, only one boundary of the integral in Eq.~(\ref{eq4}) can be extended to its corresponding surface. In this case, we propose an integral similar to Eq.~(\ref{eq4}) that depends on the electric field instead of the electrostatic potential.

In order to determine whether the potential $\psi$ is monotonic or not, we examine Eq.~(\ref{eq3}). Assume that $\psi$ is non-monotonic, such that at some point in between the two surfaces, $-d/2\le z^*\le d/2$, the electric field ${\cal E}$ vanishes. The osmotic pressure evaluated at this point yields $\Pi=2\sinh^2\left(\psi(z^*)/2\right)\ge0$. We deduce that the  electrostatic potential is monotonic for a negative osmotic pressure. Similarly, assume that the electric field is non-monotonic, such that at some point $z^{**}$ the (dimensionless) charge density, $\rho=-\sinh\psi$ of Eq.~(\ref{eq1}) vanishes. The negative osmotic pressure is then $\Pi=-{\cal E}^2(z^{**})/2\le0$, implying that the electric field is monotonic for positive osmotic pressures.

\begin{figure}[ht]
\centering
\includegraphics[width=0.95\columnwidth]{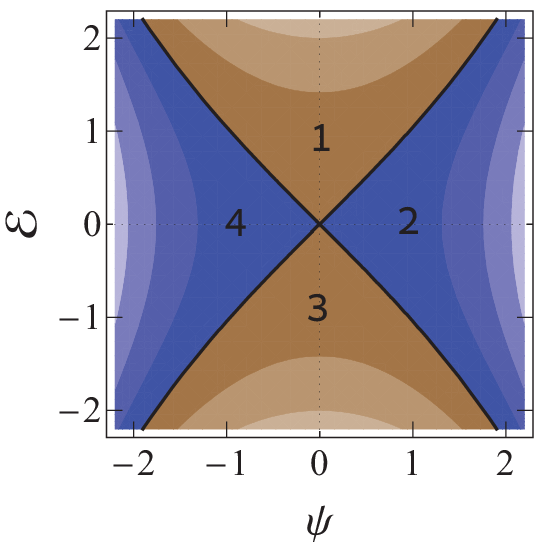}
\caption{(Color online) Contour plots of equal osmotic pressure in the $\left(\psi,{\cal E}\right)$ phase space. The lighter (brown)  regions, numbered $1$ and $3$, correspond to negative osmotic pressures, $\Pi<0$, while the darker (blue) regions, numbered $2$ and $4$, correspond to positive ones, $\Pi>0$. The black lines between the four regions correspond to infinitely separated surfaces with $\Pi=0$.}
\label{fig1}
\end{figure}

The above remarks are  illustrated in Fig.~\ref{fig1}, where contours of equal osmotic pressure are plotted in the $\left(\psi,\mathcal{E}\right)$ phase space. The $\Pi=0$ contour, which corresponds to infinitely-separated surfaces, marks the boundaries between four regions, as is indicated in the figure. It is evident that ${\cal E}\ne0$ throughout regions $1$ and $3$ of negative osmotic pressures, while $\psi\ne 0$ throughout regions $2$ and $4$ of positive osmotic pressures. We emphasize that the osmotic pressure is determined by both the boundary conditions and inter-surface separation. As a result, surfaces with given boundary conditions can exhibit repulsion at short separations and attraction at large separations or vice versa. We consider next the negative and positive osmotic pressure regimes, separately.

\subsection{Attractive osmotic pressure $\Pi<0$}\label{NegPi}

For the negative (attractive) osmotic pressures, denoted as $\Pi_{-}\equiv \Pi <0$, $\psi(z)$ is an invertible function. We  may write $\D z=-\D\psi/{\cal E}$, where the relation ${\cal E}={\cal E}\left(\psi,\Pi_{-}\right)$ is defined by Eq.~(\ref{eq3}). Integrating it once yields
\be
\label{eq5}
d=-\int_{\psi_1}^{\psi_2} \frac{\D \psi}{{\cal E}}=\pm\int_{\psi_1}^{\psi_2}\frac{\D \psi}{\sqrt{4\sinh^2\frac{\psi}{2}-2\Pi_{-}}},
\ee
where the sign of the integral ensures that $d>0$. Note that Eq.~(\ref{eq5}) is a special case of Eq.~(\ref{eq4}). Although this expression is often used, we show here that it is justified only for negative osmotic pressures. Otherwise, the interval of the integration must be limited, as is the case between symmetric surfaces, where $\Pi$ is always positive and the integration is conveniently performed between the mid-plane and any one of the two surfaces.

In the above equation, the values of the potential on the surfaces, $\psi_{1,2}$ are determined by the boundary conditions. For CP, they are prescribed, while for CC, they depend on the osmotic pressure according to Eq.~(\ref{eq3}).  Note that by the choice of $\psi$ as the integration variable for $\Pi_{-}<0$, the argument of the square root in Eq.~(\ref{eq5}) is always positive definite and no divergence can occur.

Equation~(\ref{eq5}) can be solved analytically~\cite{Safinyabook}, leading to
\be
\label{eq6}
d=\left.\mp i \,r_{-}F\left(\frac{i\psi}{2} \,;\, r_{-}\right)\right|_{\psi=\psi_1}^{\psi=\psi_2},
\ee
where $r_{-}\equiv \sqrt{-2/\Pi_{-}}$ and $F\left(t\,;\,m\right)=\int_{0}^{t}\D \theta /\sqrt{1-m^2\sin^2\theta}$
is the elliptic integral of the first kind~\cite{AbramowiczStegun}, and the choice of the sign is in agreement with Eq.~(\ref{eq5}). This equation relates $\Pi$ to the separation $d$ and the boundary conditions, without the need to specify the values of $\psi$ throughout the ionic solution. In particular, it enables the plotting of $\Pi(d)$ curves via the inverse function $d(\Pi)$.

\subsection{Repulsive osmotic pressure $\Pi>0$}

For positive (repulsive) osmotic pressures, denoted by $\Pi_{+}\equiv \Pi>0$, ${\cal E}(z)$ (and not $\psi$) is invertible. We write $\D z=\D {\cal E}/\rho$, where the charge density $\rho=-\sinh\psi$ was introduced earlier, and $\psi=\psi\left({\cal E},\Pi_{+}\right)$ according to Eq.~(\ref{eq3}). The pressure is determined by
\be
\label{eq7}
d=\int_{{\cal E}_1}^{{\cal E}_2}\frac{\D {\cal E}}{\rho}=\pm\int_{{\cal E}_1}^{{\cal E}_2}\frac{\D {\cal E}}{\sqrt{\left(1+\Pi_{+}+\half {\cal E}^2\right)^2-1}},
\ee
where ${\cal E}_{1,2}={\cal E}(\pm d/2)$ are the boundary values, and the choice of the sign before the integral ensures that $d>0$.
Because it is customary to solve for the electrostatic potential $\psi$, such an integral in terms of the electric field ${\cal E}$ might seem odd. However, we find it to be very useful as long as the osmotic pressure is positive, and conveniently relates the pressure, inter-surface distance, and surface-charge densities.

Similarly to the previous case of Sec.~\ref{NegPi}, the values of the electric field on the surfaces, ${\cal E}_{1,2}$, are determined by the boundary conditions. With our dimensionless parameters,  the constant surface-charge densities, $\sigma_{1,2}$, yield the boundary conditions ${\cal E}_{1}=4\pi\sigma_{1}\lb\ld/e$ and $\mathcal{E}_{2}=-4\pi\sigma_{2}\lb\ld/e$. These two quantities are inversely proportional to the corresponding Gouy-Chapman lengths of each surface, $\lgc=e/\left(2\pi\lb|\sigma|\right)$.
Note that by the choice of $\cal{E}$ as the integration variable for $\Pi_{+}>0$, the argument of the square root inside the ${\cal E}$ integral of Eq.~(\ref{eq7}) is positive definite, similar to the integrand of Eq.~(\ref{eq5}).

\begin{figure}[ht]
\centering
\includegraphics[width=0.95\columnwidth]{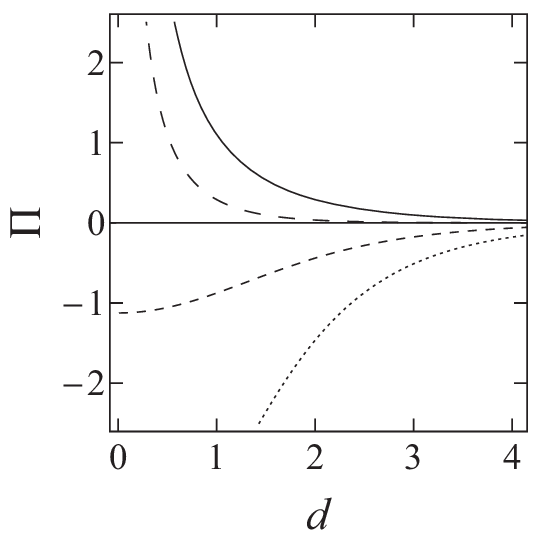}
\caption{Osmotic pressure profiles between two constant charged surfaces for the following boundary conditions (top to bottom): $({\cal E}_1,{\cal E}_2)= (1,-1)$ in solid line, $({\cal E}_1,{\cal E}_2)=(1,0)$ in long dashes,$({\cal E}_1,{\cal E}_2)=(1,1)$ in short dashes, and $({\cal E}_1,{\cal E}_2)=(2,2)$ in dotted line. The two upper repulsive profiles ($\Pi>0$) were plotted according to Eq.~(\ref{eq8}), and the lower attractive ones ($\Pi<0$) according to Eq.~(\ref{eq6}).}
\label{fig2}
\end{figure}

Equation~(\ref{eq7}) can be solved analytically, leading to
\be
\label{eq8}
d=\left.\pm r_{+} F\left(\tan^{-1}\frac{{\cal E}}{\sqrt{2\Pi_{+}}}\,;\,r_{+}\right)\right|_{{\cal E}={\cal E}_1}^{{\cal E}={\cal E}_2}.
\ee
with $r_{+}\equiv \sqrt{2/(\Pi_{+}+2)}$ and $F$ being the elliptical integral as above.
Similarly to Eq.~(\ref{eq6}), this equation relates $\Pi$ to the separation $d$ and the boundary conditions and enables the plotting of $\Pi(d)$ curves via its inverse function $d(\Pi)$. Pressure curves for different boundary conditions are plotted in Fig.~\ref{fig2}. We note that $\Pi$ can change its sign as $d$ varies, in which case both Eqs.~(\ref{eq6}) and (\ref{eq8}) should be used; the former for $\Pi<0$ and the latter for $\Pi>0$. Such a repulsion-attraction crossover is discussed in the next section.

Equations (\ref{eq6}) and (\ref{eq8}) are our main results. They demonstrate how the pressure is determined by the inter-surface separation and boundary conditions. By the correct choice of variables, the {\it electrostatic potential} for negative osmotic pressures and the {\it electric field} for positive osmotic pressures, we have obviated the need for extra symmetry arguments. In particular, it enables the calculation of the osmotic pressure for asymmetric two-surface systems.

We note that the specific choice of variables is relevant also to the calculation of the electrostatic potential, $\psi$. Once the osmotic pressure is determined by Eq.~(\ref{eq6}) or (\ref{eq8}), the potential profile $\psi(z)$, or the electric field ${\cal E}(z)$, can be evaluated accordingly. This is done by inserting the osmotic pressure, and replacing $d\to z$ and $\psi_2\to\psi(z)$ in Eq.~(\ref{eq6}), or ${\cal E}_2\to {\cal E}(z)$ in Eq.~(\ref{eq8}). Therefore, it is also possible to obtain the potential and electric field profiles for asymmetric boundary conditions using our approach.

\section{Repulsion-Attraction Crossover}
\label{sec3}

Evaluating the osmotic pressure relies on a prior knowledge of its sign. This imposes no difficulty as the criteria for the repulsion-attraction crossover  are already known within PB theory~\cite{Parsegian72,Lau99,Dan07}. For completeness, we re-derive these criteria using our above proposed framework.

For CP boundary conditions, the osmotic pressure can vanish only when the potentials on the two surfaces, $\psi_1$ and $\psi_2$, have the same sign. Otherwise, by continuity of the potential profile, $\psi$  vanishes at some point in between the two surfaces, leading to a negative osmotic pressure, as is described in Sec.~\ref{sec2}.  The crossover is obtained by taking the $\Pi_{-}\to 0^-$ limit in Eq.~(\ref{eq5}). We find that
\be
\label{eq9}
d=\pm\ln\frac{\tanh\left(\psi_2/4\right)}{\tanh\left(\psi_1/4\right)},
\ee
where the choice of $\pm$ ensures that the right-hand-side is positive, depending on whether $|\psi_2|>|\psi_1|$ or vice versa.
This defines the CP attraction region for potentials of the same sign~\cite{Dan07}
\be
\label{eq10}
\e^{-d}\,<\,\frac{\tanh\left(\psi_2/4\right)}{\tanh\left(\psi_1/4\right)}\,<\,\e^{d}.
\ee
Furthermore, surface potentials of opposite signs always result in attraction, as is explained above.

For CC boundary conditions, the osmotic pressure can vanish only when $\mathcal{E}_1$ and $\mathcal{E}_2$  at the two boundaries have the same sign, corresponding to surfaces charged with opposite signs. Otherwise, by continuity of the electric field, ${\cal E}$ vanishes at some point in between the two surfaces, leading to a positive osmotic pressure, as is described in Sec.~\ref{sec2}. The crossover is obtained by taking the $\Pi_{+}\to 0^+$ limit in Eq.~(\ref{eq7}), leading to
\be
\label{eq11}
d=\pm\ln\frac{\gamma_2}{\gamma_1},
\ee
where $\gamma_i=\sqrt{{\cal E}_i^{-2}+4}+2{\cal E}_i^{-1}$, $i=1,2$ \cite{Safinyabook}. This defines the CC attraction region~\cite{Dan07}
\be
\label{eq12}
\e^{-d}\,<\,\frac{\gamma_2}{\gamma_1}\,<\,\e^{d},
\ee

Equations (\ref{eq10}) and (\ref{eq12}) demonstrate how the sign of the osmotic pressure depends on the boundary conditions and inter-surface separation. For surface potentials or surface charges within a certain range, there is a finite separation where the osmotic pressure turns from positive to negative or vice versa. For such boundary conditions, Eq.~(\ref{eq6}) can be used to plot the negative osmotic pressure values and Eq.~(\ref{eq8}) to plot the positive ones. However, this more complex $\Pi(d)$ profile will not be further discussed in the present study. 

\section{Repulsive Pressure Regimes and Scalings}
We further demonstrate the advantages of our approach employing the electric field, Eqs. (\ref{eq7}) and (\ref{eq8}), by obtaining the pressure scaling in different repulsive pressure regimes. For simplicity, in this section $\Pi$ is the notation for the positive (repulsive) pressure, and the scaling laws are presented in terms of dimensionless variables. They can be easily converted back into the physical ones using Eqs.~(\ref{eq5t})-(\ref{eq7t}).

\subsection{Ideal gas regime}
Assume that $\Pi\gg 1$ and  $\Pi\gg \mathcal{E}_{i}^2$ for the two surfaces, $i=1,2$. Keeping only the $\Pi^2$ term in the square root in Eq.~(\ref{eq7}) yields
\be
\label{eq13}
\Pi\approx\frac{1}{d}\left|\mathcal{E}_{2}-\mathcal{E}_{1}\right|.
\ee
Namely, the osmotic pressure is given by the ideal gas pressure, where the excess concentration of ions is determined by the overall charge on the surfaces. In particular, the above result restores the result for symmetric surfaces~\cite{Safinyabook}. The range of validity of this regime is $d\ll \left|\mathcal{E}_{2}-\mathcal{E}_{1}\right|$ while $1/d\gg \mathcal{E}_{i}^2/\left|\mathcal{E}_{2}-\mathcal{E}_{1}\right|$, for both surfaces, $i=1,2$.

\subsection{Gouy-Chapman regime}
Assume that $\mathcal{E}_{1}<0<\mathcal{E}_{2}$ and $\mathcal{E}_{1},\mathcal{E}_{2}\gg\sqrt{\Pi}$, corresponding to a large surface charge. The first argument of the elliptic function in Eq.~(\ref{eq8}) can then be approximated by $\pm\pi/2$, and the osmotic pressure becomes independent of the surface charge, according to
\be
\label{eq14}
d\approx 2r K(r),
\ee
where $K(m)=F\left(\pi/2\,;\,m\right)$ is the complete elliptic integral of the first kind~\cite{AbramowiczStegun}, and $r=\sqrt{2/(\Pi+2)}$ as was defined above. Further assuming that $\Pi\gg1$, we arrive at
\be
\label{eq15}
\Pi\approx\frac{2\pi^2}{d^2}.
\ee
As the pressure is independent of the surface charge in the Gouy-Chapman regime, $\Pi$  coincides with the known result for symmetrical surfaces~\cite{Safinyabook}. The range of validity of this regime is $|\mathcal{E}_{1}|,|\mathcal{E}_{2}|\gg 1/d\gg 1$.

\subsection{Intermediate regime}
In the limit of large surface-charge densities but small pressures, we keep the lowest order in $\Pi\ll 1$ in Eq.~(\ref{eq14}), resulting in
\be
\label{eq16}
\Pi=32{\rm e}^{-d}.
\ee
In this intermediate regime, $\Pi$ also coincides with the result for symmetrical surfaces~\cite{Safinyabook}. The range of validity of this regime is $d\gg 1\gg 1/|\mathcal{E}_{1}|,1/|\mathcal{E}_{2}|$.

\subsection{Debye-H{\"u}ckel regime}
For weak electrostatic interaction, $\Pi,\mathcal{E}_{i}^{2}\ll 1$ for the $i=1,2$ surfaces, and the integral form of Eq.~(\ref{eq7}) is well approximated by
\be
\label{eq17}
d=\int_{{\cal E}_1}^{{\cal E}_2}\frac{\D {\cal E}}{\sqrt{2\Pi+{\cal E}^2}}.
\ee
This integral is solvable, leading to
\be
\label{eq19}
d=\ln\left(\frac{{\cal E}_2+\sqrt{2\Pi+{\cal E}_2^2}}{{\cal E}_1+\sqrt{2\Pi+{\cal E}_1^2}}\right).
\ee
For $\Pi\ll \mathcal{E}_{i}^{2}$, expanding the right-hand-side of the above equation results in
\be
\label{eq20}
\Pi=-2\mathcal{E}_{1} \mathcal{E}_{2}\exp(-d).
\ee
Note that this pressure is indeed positive, since $\mathcal{E}_{1}\mathcal{E}_{2}<0$ within this limit. The range of validity of this result is $|\mathcal{E}_{1}|,|\mathcal{E}_{2}|\ll1$ while $d\gg1$.

For $\Pi\gg \mathcal{E}_{i}^{2}$, the argument of the logarithm in Eq.~(\ref{eq19}) is close to unity. Expanding the logarithm yields
\be
\label{eq18}
\Pi=\half \left(\frac{\mathcal{E}_{2}-\mathcal{E}_{1}}{d}\right)^2.
\ee
The range of validity of this result is $\left|\mathcal{E}_2-\mathcal{E}_1\right|/|\mathcal{E}_i|\gg d\gg \left|\mathcal{E}_2-\mathcal{E}_1\right|$ for both $i=1,2$ surfaces.
Equations (\ref{eq18}) and (\ref{eq20}) restore the results of the linear Debye-H{\"u}ckel theory for asymmetrically charged surfaces in the corresponding limits~\cite{Dan13}.

\section{The Modified Poisson-Boltzmann (MPB) Theory}
\label{sec4}

The scheme presented above for the standard PB theory can be generalized to other augmented PB theories. We demonstrate it for the sterically modified Poisson-Boltzmann theory (MPB)~\cite{Borukhov97,Borukhov00,Kilic07}, which takes into account steric effects due to the finite size of ions. The ion size adds another length scale to the standard PB theory, $a$, which defines the close-packing density of ions, $a^{-3}$. For 1:1 monovalent salt, the MPB equation in Gaussian units reads
\be
\label{eq32}
\frac{\eps}{4\pi}\Psi''=\frac{2n_b e\sinh\left(\frac{e\Psi}{\kbt}\right)}{1+2n_b a^{3}\left[\cosh\left(\frac{e\Psi}{\kbt}\right)-1\right]}.
\ee
It is evident that the charge density on the right hand side is bounded in absolute value by $ea^{-3}$ and reduces to the standard PB form of Eq.~(\ref{eq1}) for $n_b a^3\ll 1$.
The first integral of the MPB equation can be obtained analytically and yields the pressure across the electrolyte~\cite{Dan09},
\be
\label{eq21}
P=-\frac{\eps}{8\pi}\Psi'^2+a^{-3}\kbt\,\ln\left(1+\frac{2n_b a^3}{1-2n_b a^3}\cosh\frac{e\Psi}{\kbt}\right).
\ee
While the contribution of the Maxwell stress tensor is the same as in Eq.~(\ref{eq2}), the ideal gas van 't Hoff pressure is replaced  by a more complex lattice-gas logarithmic term. \ra{Note that steric effects can be accounted for also by considering short-range non-Coulombic repulsive interactions between ions (see, e.g., Ref.~\cite{Pitzer73}). Such models produce different expressions for the pressure, depending on the exact form of the interaction.}

As before, it is more convenient to use dimensionless variables:
\bea
\psi \equiv e\Psi/\kbt, &~~~~~ \tilde{{\cal E}}\equiv -\D\psi/\D \tilde{z}, \nonumber\\
\tilde{z}\equiv Z/\lambda, & \tilde{d}\equiv D/\lambda,
\eea
with the characteristic length scale not being the Debye length, but $\lambda \equiv 1/\sqrt{4\pi\lb a^{-3}}$.  The pressure is rescaled according to $\tilde{p}= P/\left(\kbt a^{-3}\right)$, and similarly for the osmotic pressure, $\tilde{\Pi}=P_{\rm osm}/\left(\kbt a^{-3}\right)$.  We also introduce a new variable, $\Phi=2n_b a^3$ that is the volume fraction of ions in the bulk electrolyte. For small $\Phi\ll 1$ values, steric effects are negligible and the MPB theory reduces to the standard PB one.

Comparing the dimensionless variables of the MPB theory with the previously defined ones of the standard PB theory we find that
\be
\tilde{d}=\Phi^{-1/2}d, ~~~ \tilde{\Pi}=\Phi\Pi, ~~{\rm and}~~ \tilde{\cal E}=\Phi^{1/2} {\cal E}.
\ee
With these variables, the dimensionless osmotic pressure is given by
\be
\label{eq22}
\tilde{\Pi}=-\half \tilde{\cal E}^2+\ln\left[1+\Phi\left(\cosh\psi-1\right)\right].
\ee
Following the same arguments presented after Eq.~(\ref{eq3}), we deduce that $\psi$ is monotonic for $\tilde{\Pi}=\tilde{\Pi}_{-}<0$ and $\tilde{\cal E}$ is monotonic for $\tilde{\Pi}=\tilde{\Pi}_{+}>0$.

For negative osmotic pressures, $\psi$ is chosen as the integration variable, and the integral equation relating $\tilde{d}$ and $\tilde{\Pi}_{-}$ reads
\begin{align}
\label{eq23}
\tilde{d}&=-\int_{\psi_1}^{\psi_2}\frac{\D\psi}{\tilde{\cal E}}\nonumber\\
&=\pm\int_{\psi_1}^{\psi_2}\frac{\D\psi}{\sqrt{2\left(\ln\left[1+\Phi\left(\cosh\psi-1\right)\right]-\tilde{\Pi}_{-}\right)}}.
\end{align}
The sign of the right-hand-side is chosen such that $\tilde{d}$ is positive. For small $\Phi$ values, the logarithm term in the denominator can be expanded to linear order in $\Phi$. Converting $\tilde{d}\to d$ and $\tilde{\Pi}_{-}\to\Pi_{-}$, the standard PB form of Eq.~(\ref{eq5}) is restored in this limit.

For positive osmotic pressures, $\tilde{\cal E}$ plays the role of the integration variable, and the  following relation is obtained:
\begin{align}
\label{eq24}
\tilde{d}&=\int_{\tilde{\cal E}_1}^{\tilde{\cal E}_2}\frac{\D \tilde{\cal E}}{\rho}\nonumber\\
&=\pm\int_{\tilde{\cal E}_1}^{\tilde{\cal E}_2}\D {\cal E}\frac{\exp\left(\tilde{\Pi}_{+}+\half \tilde{\cal E}^2\right)}{\sqrt{\left[\exp\left(\tilde{\Pi}_{+} + \half \tilde{\cal E}^2\right) -1 +\Phi \right]^2 -\Phi^2}}.
\end{align}
For small $\Phi$ values, the argument of the exponents, $\tilde{\Pi}_{+}+\tilde{\cal E}^2/2=\Phi\left(\Pi_{+}+{\cal E}^2/2\right)$ become small. Approximating the numerator to zeroth order in $\Phi$ and the denominator to first order in $\Phi$,  the standard PB results of Eq.~(\ref{eq7}) for $d$, $\Pi_{+}$  and ${\cal E}$ are restored.

\begin{figure}[ht]
\centering
\includegraphics[width=0.95\columnwidth]{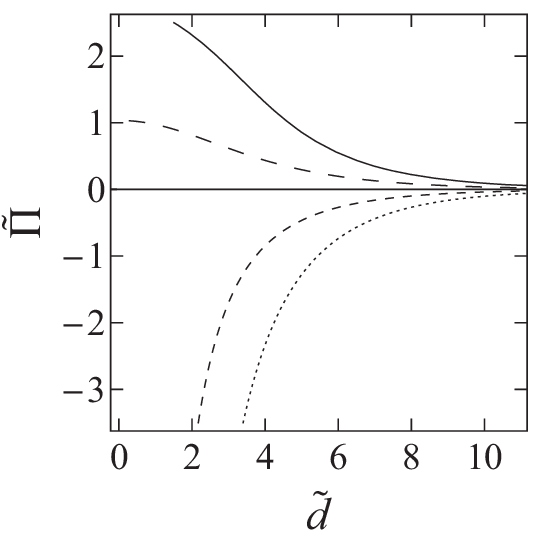}
\caption{Osmotic pressure profiles between homogeneously charged surfaces within MPB theory for (top to bottom): $(\psi_0,\psi_1)=(5,5)$ in solid line, $(\psi_0,\psi_1)=(3,3)$ in long dashes, $(\psi_0,\psi_1)=(-0.5,0.5)$ in short dashes, and $(\psi_0,\psi_1)=(-0.2,0.2)$ in dotted line. The upper two repulsive profiles ($\Pi>0$) were plotted according to Eq.~(\ref{eq24}), and the lower attractive ones ($\Pi<0$) according to Eq.~(\ref{eq23}). }
\label{fig3}
\end{figure}

Equations~(\ref{eq23}) and (\ref{eq24}) are the MPB analogs of Eqs.~(\ref{eq5}) and (\ref{eq7}) of the standard PB theory. Although they cannot be integrated analytically, as is the case for the standard PB theory, they provide a direct relation between the separation, the osmotic pressure and the boundary conditions, and enable the numerical plotting of pressure curves, as is illustrated in Fig.~\ref{fig3}. It is possible to use Eqs.~(\ref{eq23}) and (\ref{eq24}) to derive approximate forms of the MPB pressure in different electrostatic regimes. Such a calculation exceeds the scope of the current study and is left for future studies.

\section{Conclusions}
\label{sec5}
In this work, we present a different approach for calculating the osmotic pressure within PB theory.  Analytical results for the osmotic pressure are derived for arbitrary boundary conditions, both in the standard PB theory and in the augmented sterically modified PB theory. As many surface force experiments involve two bounding surfaces with asymmetric boundary condition (either $\sigma_1\ne \sigma_2$ or $\psi_1\ne\psi_2$), such analytical expressions should be helpful.

A useful concept found in the present work is that the electric field, ${\cal E}$, can be a more suitable variable to work with, compared to the electrostatic potential, $\psi$. Explicitly, the sign of the osmotic pressure dictates which of the two profiles (${\cal E}$ or $\psi$) is monotonic between the two surfaces, and, therefore, can be inverted. Using the electric field as the natural variable is especially convenient for CC boundary conditions that dictate the values ${\cal E}_{1,2}$ on the boundaries.

We clarify the limitations of the framework presented in this paper. Our scheme relies on an analysis of the analytically obtained first integral of the PB equation (and its possible generalizations). As a first integral does not exist for cylindrical or spherical geometries, we restrict ourselves to planar geometry. Furthermore, augmentations of the PB theory sometimes result in complicated first integrals. In such cases,relations of the form ${\cal E}\left(\psi,\Pi\right)$ and $\psi\left({\cal E},\Pi\right)$ and the resulting osmotic pressure cannot be obtained analytically.

\ra{In this work, we focused on the CP and CC boundary conditions that suffice for describing many experimental setups}. In a more physical picture, however, surface charges are regulated by association and dissociation of charged groups in equilibrium with the ionic solution. Such processes can be accounted for using the so-called {\it charge regulation} (CR) boundary condition~\cite{Ninham71,Chan75,TomerCR}. \ra{The CR case can also be analyzed using our framework, as will be shown in a future study.}

Finally, we note that our framework can be used to determine electrostatic potential profiles, even when the inter-surface separation, $d$, is related to the osmotic pressure only via an integral form, as in Eqs.~(\ref{eq23}) and (\ref{eq24}). In such cases, the osmotic pressure is determined numerically by solving the integral equation. Then, the value of $\Pi$ can be inserted in the integrand, and the potential (or electric field) profile can be found by numerical integration from one of the boundaries to an arbitrary $z$.

\vskip 0.5cm
{\it Acknowledgments.~} {This article is dedicated to the memory of Loic Auvray. An outstanding gentleman of science who introduced us to a wealth of physical phenomena and concepts in polymers, polyelectrolytes, biophysics and charged soft matter systems. His unique kindness and deep understanding of physics will not be forgotten.}
We thank R. Calman, T. Markovich, and R. Podgornik for helpful discussions and suggestions. This work was supported by the Israel Science Foundation (ISF) under grant No. 438/12, the U.S.- Israel Binational Science Foundation (BSF) under grant No. 2012/060, and the ISF-NSFC joint research program under grant No. 885/15.



\begin{thebibliography}{99}

\bibitem{Israelachvily}
J. N. Israelachvili, {\it Intermolecular and Surface Forces}, 3rd ed. (Academic, New York, 2011).

\bibitem{VO} E. J. Werwey and J. Th. G. Overbeek, {\it Theory of the Stability of
Lyophobic Colloids} (Elsevier, New York, 1948).

\bibitem{Safinyabook} T. Markovich, D. Andelman, and R. Podgornik, in {\it Handbook of Lipid Membranes}, edited by C. Safinya and J. R{\"a}dler (Taylor and Francis, to be published).
\bibitem{Patchy} R.M. Adar, D. Andelman, and H. Diamant, Adv. Colloids Interf. Sci. {\bf 247}, 198  (2017).
\bibitem{Dan09} D. Ben-Yaakov, D. Andelman, D. Harries, and R. Podgornik,  J. Phys. Chem. B {\bf 113}, 6001 (2009).
\bibitem{Maggs16} A. C. Maggs and R. Podgornik, Soft matter {\bf 12}, 1219 (2016).
\bibitem{Grahame47} D. C. Grahame, Chem. Rev. {\bf 41}, 441 (1947).
\bibitem{Henderson81} D. Henderson and L. Blum, J. Chem. Phys. {\bf 75}, 2025 (1981).
\bibitem{Evans99} D. F. Evans and H. Wennerstrom, {\it "The Colloidal Domain"}, 2nd ed. (VCH Publishers, New York, 1999)
\bibitem{Dean03} D. S. Dean and R. Horgan, Phys. Rev. E {\bf 68}, 061106 (2003).

 \bibitem{Meier04} A. A. Meier-Koll, C. C. Fleck, and H. H. von Gr{\"u}nberg, J. Phys.: Condens.
Matter {\bf 16}, 6041 (2004).
\bibitem{Safran05} S. A. Safran, Europhys. Lett. {\bf 69},  826 (2005).

\bibitem{AbramowiczStegun}
M. Abramowicz and I. A. Stegun, {\it Handbook of Mathematical Functions}, (Dover, New York, 1965).






\bibitem{Parsegian72} V. A. Parsegian and D. Gingell, Biophys. J. {\bf 12}, 1192 (1972).
\bibitem{Lau99} A. Lau and P. Pincus, Eur. Phys. J. B {\bf 10}, 175 (1999).
\bibitem{Dan07} D. Ben-Yaakov, Y. Burak, D. Andelman, and S. A. Safran,  Europhys. Lett. {\bf 79}, 48002 (2007).

\bibitem{Dan13} D. Ben-Yaakov, D. Andelman, H. Diamant,  Phys. Rev. E {\bf 87}, 022402 (2013).
\bibitem{Borukhov97} I. Borukhov, D. Andelman, and H. Orland, Phys. Rev. Lett. {\bf 79}, 435 (1997)
\bibitem{Borukhov00} I. Borukhov, D. Andelman, and H. Orland,  Electrochim. Acta {\bf 46}, 221 (2000).
 \bibitem{Kilic07} M. S. Kilic, M. Z. Bazant, and A. Ajdari, Phys. Rev. E {\bf 75}, 021502 (2007).
\bibitem{Pitzer73} K.S. Pitzer, J. Phys. Chem. {\bf 77}, 268 (1973).
\bibitem{Ninham71} B. W. Ninham and V. A. Parsegian, J. Theor. Biol.
{\bf 31}, 405 (1971).
\bibitem{Chan75} D. Chan, J. W. Perram, L. R. White, and T. W. Healy,
 J. Chem. Soc. Faraday Trans. I {\bf 71}, 1046 (1975).
 \bibitem{TomerCR} T. Markovich, D. Andelman, and R. Podgornik,
Europhys. Lett. {\bf 113}, 26004 (2016).


\end{thebibliography}
\end{document}